\documentclass[preprint,12pt]{elsarticle}

\usepackage{amssymb}

\usepackage{amsmath}

\def\ba{\begin{eqnarray}}
\def\ea{\end{eqnarray}}
\def\lb{\label}
\def\nn{\nonumber \\}
\def\bi{\bibitem}

\def\D{\Delta}

\def\n{\bar{n}}

\journal{Physics Letters B}

\begin{document}

\begin{frontmatter}



\title{Hidden asymmetry and long range rapidity correlations}

\author[a,b]{A. Bialas}
\ead{bialas@th.if.uj.edu.pl}

\author[c]{A. Bzdak}
\ead{abzdak@bnl.gov}

\author[a,b]{K. Zalewski}
\ead{zalewski@th.if.uj.edu.pl}

\address[a]{H. Niewodniczanski Institute of Nuclear Physics, Polish Academy of Sciences, Radzikowskiego 152, 31-342 Krakow, Poland}
\address[b]{M. Smoluchowski Institute of Physics, Jagellonian University,\\Reymonta 4, 30-059 Krakow, Poland}
\address[c]{RIKEN BNL Research Center, Brookhaven National Laboratory,\\Upton, NY 11973, USA}

\begin{abstract}
Interpretation of long-range rapidity correlations in terms of the fluctuating rapidity density distribution of the system created in high-energy collisions is proposed. When applied to recent data of the STAR coll., it shows a substantial asymmetric component in the shape of this system  in central Au-Au collisions, implying that boost invariance is violated on the event-by-event basis {\it even} at central rapidity. This effect may seriously influence the hydrodynamic expansion of the system.
\end{abstract}


\end{frontmatter}


{\bf 1.} It is now widely recognized that long-range correlations (LRC) in
rapidity originate at the early stages of the collision, before the longitudinal expansion separates the particles by large distances. Such correlations can thus be used as a 
probe of the  initial conditions of the evolution.
This is particularly interesting for hydrodynamic description of particle production, as the initial conditions strongly influence the 
evolution of the system (called henceforth "a fireball") expanding according to the rules of hydrodynamics  \cite{init}. 
The event-by-event fluctuations of the initial conditions in the transverse plane were already shown to induce several interesting features in the transverse momentum correlations observed in the final state \cite{v23}. For example the recently observed so-called triangular flow is entirely due to the asymmetric fluctuations in the transverse plane \cite{v3}. In the present paper we discuss the effects of the initial state fluctuations on the  correlations between the {\it longitudinal} momenta. We show how the measurements of the LRC in rapidity can be interpreted in terms of the event-by-event fluctuating shape  of the fireball. We argue, using the relevant data of STAR coll. \cite{star,tar}, that such analysis can uncover some hitherto unobserved features of particle production.

A special case of LRC are forward-backward correlations where one
compares particle distributions in two intervals located symmetrically
in the forward and backward hemispheres. They were extensively studied
since the early times of high-energy physics \cite{kitw}. In most of these  studies  only the global density fluctuations were considered \cite{cy,bzdd,bzd,bzbf} and  data    were interpreted as evidence for strong event-by-event fluctuations of the multiplicity of the produced particles. With the increasing precision of data and larger observed  particle densities, however, it seems useful to consider the more general scenario, with the event-by-event fluctuations of both {\it multiplicity} and {\it shape} of the created system.  In the present paper we show that this approach allows to obtain some direct information about the object produced in a high-energy collision.
 
To this end we apply the recently proposed method \cite{bznpa,chwast,bzplb} of a systematic study of the factorial moments of multiplicity distribution in several well separated rapidity bins (see also \cite{lappi}). Let us add that, as shown in \cite{bznpa,bzplb}, such measurements allow also to discriminate between various models of the multiparticle production and thus to understand better the mechanism of such processes.

In the next section the problem is formulated in terms of the generating functions. In Section  3 the data of STAR collaboration \cite{star} are analyzed and it is shown that
they imply existence of a substantial asymmetric component in the fireball shape for central Au-Au collisions. A discussion of this result is given in section 4. Our conclusions are listed in the last section.

{\bf 2.} Consider a fireball created in a single collision and a rapidity bin
$\Delta_i$. The number of particles in $\Delta_i$ is a random number, whose
distribution depends on the initial conditions of the collision, on parameters of the fireball evolution and on details of hadronization. All these factors  may be summarized 
in a set of parameters, $Q\equiv (q_1,q_2,...)$ which are also 
random numbers. They may be, for instance, the impact parameter of the collision, the number of participating nucleons, inhomogeneities of the expanding fluid, etc.
Let us denote the average number of particles falling into $\Delta_i$ at a given
$Q$ by $\n_i(Q)$. It is related to the corresponding particle
density by
\ba
\n_i(Q)=\int_{\D_i} \rho(y) dy \approx \rho(y_i) \D_i \lb{enav}
\ea
We call {\it dynamic} the fluctuations of $\n_i$ resulting from the
randomness of $(q_1,q_2,...)$. Our purpose is to estimate these fluctuations. 

Even for a fixed $Q$, however, the actual number of particles in $\Delta_i$  fluctuates around $\n_i$. This is the noise we want to correct for. To this end
we  assume that it is possible to choose the parameters $(q_1,q_2,...)$ so that for each set $Q$ the
fluctuations around $\n_i(Q)$ are dominantly random i.e.  approximately Poissonian. Note, that we do not have to know the set $Q$. It is enough to assume that it exists. Under this
assumption, for $B$ bins at given $Q$ the probability distribution for the
occupation numbers $n_1,\ldots,n_B$ is
\ba
P(n_1,...,n_B;\n_1,...,\n_B)=p(n_1;\n_1)...p(n_B;\n_B);\;\;\; p(n;\n)=e^{-\n}\frac{\n^n}{n!}
\lb{ps0}
\ea
where the argument $Q$ has been omitted. The observed distribution is the
average over $Q$, or equivalently over the averages $\n_i(Q)$\footnote{To illustrate the idea, consider a Monte Carlo simulation in which the probability density of finding a number of particles at some momentum is not fixed but depends on some (random) parameters $Q$. 
The probability density at a given $Q$ is our $\rho(y)$. Fluctuations resulting from fluctuations of $Q$ are the dynamical fluctuations. The remaining random event-by-event fluctuations are our purely statistical fluctuations.}:
\ba
P(n_1,...,n_B)=\int d\n_1...d\n_B W(\n_1,...,\n_B) p(n_1;\n_1)...p(n_B;\n_B) \lb{ps}
\ea
where $W(\n_1,...,\n_B)$ is the probability distribution of the set $[\n_1,...\n_B]$, 
characterizing the distribution of the densities of the produced fireballs, i.e. the basic quantity of interest.

From the well-known property of the Poisson distribution 
\ba
\sum_n \frac{n!}{(n-k)!} p(n;\n)=\n^k
\ea 
one easily derives
\ba
F_{i_1,..,i_B}=\int d\n_1...d\n_B W(\n_1,...,\n_B)\n_1^{i_1},...,\n_B^{i_B}=
\left<\n_1^{i_1},...,\n_B^{i_B}\right>_W , \lb{fac}
\ea
where $F_{i_1,..,i_B}$ are the factorial moments of the distribution (\ref{ps}):
\ba
F_{i_1,..,i_B}=\left< \frac{n_1!}{(n_1-i_1)!}...\frac{n_B!}{(n_B-i_B)!}\right>
\ea
Eq.(\ref{fac})  shows that measurement of factorial moments of the observed multiplicity distribution gives directly the moments of the fluctuating fireball densities \cite{bpint}.

{\bf 3.} As an application of  (\ref{fac}) we shall show that  the published results of the STAR collaboration \cite{star} give evidence for a substantial asymmetric component in fluctuations of the fireball created in Au-Au collisions at $\sqrt{s}=200$ GeV. In this experiment the fluctuations of multiplicity observed 
in two rapidity bins, symmetric with respect to $y_{c.m.}=0$ [at $0.8\leq|y|\leq 1.0$], were measured for various centralities,  selected according to the number of particles observed in the central bin, located  also symmetrically around $y_{c.m.}=0$. At the highest centrality and at the highest distance between the bins, the measurements \cite{star,tar} give
\ba
D_{ff}^2\equiv <n_f^2>-<n_f>^2= 350\pm 17;\; \nn D_{fb}^2\equiv <n_fn_b>-<n_f>^2
=202\pm 17;\;\;\;<n_f>=96\pm 5. \;\;\;\;\;\;\lb{disp}
\ea
where the indices $f$ and $b$ refer to the forward and backward bins, respectively, and 
$n_f$ and $n_b$ denote the actually observed numbers of particles in these bins.

The factorial moments are related to (\ref{disp}) by
\ba
F_{20}\equiv <n_f(n_f-1)>=<\n_f^2>_W=D_{ff}^2+<n_f>^2-<n_f>;\;\;\; \nn F_{11}\equiv <n_fn_b>=<\n_f\n_b>_W=D_{fb}^2+<n_f>^2
\ea
Noting that $(\n_f\pm \n_b)^2= \n_f^2+\n_b^2\pm 2\n_f\n_b$ we thus obtain for  the asymmetric and symmetric fluctuations
\ba
D_-^2\equiv \frac14<(\n_f-\n_b)^2>_W=\frac12[F_{20}-F_{11}]=26\pm 12;\nn
D_+^2\equiv \frac14<(\n_f+\n_b)^2>_W-<n_f>^2=\frac12[F_{20}+F_{11}]-<n_f>^2 =228\pm 12 . \lb{dpm}
\ea
Using (\ref{disp}) we have
\ba
D_-=5.1\pm 1.2;\;\;\;  D_+=15.1\pm 0.4.
\ea
One sees that, although the symmetric fluctuations  dominate, there is also a substantial asymmetric component. Indeed, the ratio $D_-/D_+\approx 1/3$. Thus one has to conclude  that created fireballs are not necessarily symmetric\footnote{Obviously, for a symmetric fireball $D_-=0$.}.
This observation implies that the standard assumption of boost invariance is violated on the event-by-event level {\it even} at $y \approx 0$. As the effect is expected to be   stronger at the early times (because the expansion has a natural tendency to smooth out the original inhomogeneities),  this observation may have important consequences for the theoretical description of the process (e.g. for the hydro calculations).

It is interesting to see how this result compares with $pp$ collisions. For this case the STAR collaboration \cite{star} gives $D^2_{ff}= 0.572\pm 0.030$ and $D^2_{fb}=0.027\pm 0.003 $.  Unfortunately,  the value of $<n_f>$ was not
given in \cite{star}. To obtain a rough estimate, we have used  $<n_f>=0.46\pm 0.03$ taken from \cite{phob}. This gives $D_-=0.21\pm 0.05$ and $D_+=0.26\pm 0.04$, indicating that in this case (i) the relative fluctuations are stronger\footnote{As is seen from the inequality $[D/<n>]_{pp} > [D/<n>]_{AuAu}$.} and (ii) the asymmetry of the  fireballs is even more important. This was to be expected from  the earlier observation \cite{bzdd} that the UA5 pp data \cite{UA5} are consistent with the presence of two asymmetric contributions (most likely representing remnants of the forward and backward moving projectiles).

{\bf 4.} The results discussed in the previous section demand a technical comment. The point is that the STAR measurements were performed {\it at a fixed number $N_{ch}$ of particles observed in the central bin} and then averaged over $N_{ch}$. The importance of this condition was  first noted by Lappi and McLerran who analyzed this effect in detail \cite{lappi}. Here we   add only a remark that although this procedure may   indeed change significantly   $D_+$, it does not affect $D_-$.  This can be seen from the identity
\ba 
D_-^2=\frac12 \left[D_{ff}^2-<n_f>-D_{bf}^2\right]
\ea
showing that $D_-^2$ is insensitive to the order of averaging over $N_{ch}$. 

The observed asymmetries find a natural explanation if,  at RHIC energies,  the remnants of the projectiles are still present even in the central rapidity region \cite{bc,bb}.  Indeed, the contributions from the forward and backward moving projectiles are naturally asymmetric \cite{bc}. Since they are expected to fluctuate quasi-independently, they  produce -generally- an asymmetric fireball. 

Since the STAR data give only moments of the second order, it is clear that  introducing such two components gives enough freedom  to obtain a correct value for  $D_-$. It is not clear, however, if these two contributions are really sufficient to describe the physics \cite{bb,bbc} of the observed asymmetry. It may well be that it is necessary to add a third (approximately symmetric)  component, describing the multiple parton-parton interactions \cite{dpm}. Measurement of moments of higher order is necessary to answer this interesting question. The higher moments are also needed if one wants to pin down more precisely the details of the distributions.

In this context one may add an obvious remark that the relevant measurements at LHC energies shall be also very interesting and actually may  help to disentangle this problem. Indeed, at these much higher energies, the remnants from the projectiles in the central rapidity region are expected to be small and perhaps even  completely die out. Therefore one expects their contribution to asymmetry  to be significantly smaller.

{\bf 5.} Some comments are in order. 

(i) As already mentioned in Section 2, the density $\rho(y)$, whose fluctuations we propose to study, summarizes effects of all processes leading to the observed final state. In particular, it can be modified by  hadronization: the density of gluons is replaced by densities of particles and resonances.  Our poissonian Ansatz (\ref{ps0}), (\ref{ps}) removes the purely statistical fluctuations, but does not remove the possible dynamical fluctuations which may be induced during hadronization, for example the resonance production.  These fluctuations are, for obvious reasons,  much more difficult to control, as they are model-dependent. We would like to emphasize, however, that these are the genuine dynamical effects which  provide relevant physical information about the process of particle production. Therefore they should be included in the analysis.

(ii) An estimate  of the hadronization effects  is of course of great interest, since otherwise it is not possible to make definite statements about the initial conditions which are of main interest. We have therefore worked out how the resonance production can modify our main results presented  in Section 3. We have considered an extreme  situation in which all observed pions are decay products of resonances. For $\rho$ production, the correction to $D_-$ does not exceed 8-10\%, for the transverse momentum of the $\rho$
up to 1 GeV. Similar results are obtained for $\omega$ production and for production of  heavy (1.5 GeV) clusters decaying into 3 particles. The corrections can reach up to 20\% in $pp$ collisions, still inside the quoted errors. The reason for this small effect is the small bin size $\Delta \eta = 0.2$, which makes it unlikely to have more than one decay particle in the bin. We therefore conclude that neither resonance production nor clustering effects can explain the asymmetry observed in data. 

(iii) It is of course not excluded that another, hitherto unknown, effect can be responsible
for the observed asymmetry. For the moment, however, its most likely explanation 
is the asymmetry in the initial conditions. 

{\bf 6.} In conclusion, it is argued that the systematic study of the factorial moments  of the multiplicity distribution in several rapidity intervals represents a powerful tool allowing to investigate, on event-by-event  basis,  the longitudinal structure of  systems ("fireballs") created in  high-energy collisions. It was shown that, when applied to the data of STAR collaboration \cite{star}, this method allows to uncover the importance  of {\it asymmetric} fireballs produced in the {\it symmetric} Au-Au collisions. This result seems interesting, since it suggests that the hypothesis of boost invariance is violated on the event-by-event level {\it even in the central rapidity region}. It is also interesting to note that the effect seems more significant for $pp$ collisions, thus indicating a violation of boost-invariance in "elementary" collisions. We feel that these observations should be seriously taken into account in modeling the particle production processes.

It should be emphasized that the method we discuss is very general and flexible. It can  be applied to any specific sample  of events, e.g.,  those associated with a large transverse momentum jet and/or selected according to overall multiplicity, transverse momentum and many others. Hopefully, the coming measurements at LHC will be able to exploit its full capacity.

{\bf Acknowledgements}

This investigation was supported in part by the grant N N202 125437 of the Polish Ministry of Science and Higher Education (2009-2012) and the U.S. Department of Energy under Contract No. DE-AC02-98CH10886.

\end{document}